\documentclass[10pt,twocolumn]{article}
\usepackage{graphicx}
\usepackage{amsmath,amssymb,times,cite}

\DeclareMathOperator{\sgn}{sgn}

\newcommand\raiseT[2]{%
\setbox0\hbox{$#1{#2}$}\raise\dp0\box0}
\usepackage{amsmath}

\title{\LARGE\textbf{WISE: A Computer System Performance Index Scoring Framework}}
\author{Lorenzo Luciano$^{1,*}$, Imre Kiss$^{2}$, Peter William Beardshear$^{1}$, Esther Kadosh$^{1}$, and A. Ben Hamza$^{3}$\\
$^{1}$Amazon, AWS\\
Boston, MA, USA\\
$^{2}$Amazon, Alexa AI-Natural Understanding\\
Cambridge, MA, USA\\
$^{3}$Concordia Institute for Information Systems Engineering\\
Concordia University, Montreal, QC, Canada
}
\date{}

\usepackage{amsmath}
\usepackage{booktabs}
\usepackage[table]{xcolor}
\usepackage{verbatim}

\definecolor{lightgray}{gray}{0.9}
\definecolor{lightblue}{rgb}{0.93,0.95,1.0}

\usepackage{booktabs,caption}
\usepackage{threeparttable}
\usepackage{url}

\date{}

\begin{document}
\maketitle

\begin{abstract}
The performance levels of a computing machine running a given workload configuration are crucial for both users and providers of computing resources. Knowing how well a computing machine is running with a given workload configuration is critical to making proper computing resource allocation decisions. In this paper, we introduce a novel framework for deriving computing machine and computing resource performance indicators for a given workload configuration. We propose a workload/machine index score (WISE) framework for computing a fitness score for a workload/machine combination. The WISE score indicates how well a computing machine is running with a specific workload configuration by addressing the issue of whether resources are being stressed or sitting idle wasting precious resources. In addition to encompassing any number of computing resources, the WISE score is determined by considering how far from target levels the machine resources are operating at without maxing out. Experimental results demonstrate the efficacy of the proposed WISE framework on two distinct workload configurations.
\end{abstract}

\bigskip
\noindent\textbf{Keywords}:\, Machine Performance; Workload; Migration Services; Compute Performance; Virtual Machines; Machine Score; Cloud Computing.

\section{Introduction}
Determining how well a machine is performing with a given workload configuration is not easily determined, as it can be approached from many different angles. Depending on the workload, optimum performance levels may also differ. While one workload may require a certain memory buffer, another one may excel by pushing the central processing unit (CPU) utilization boundary. Choosing the right computing configuration is a challenging problem and often becomes more difficult as a result of the vast amount of virtual machine (VM) instance types available on cloud computing platforms. Each instance type varies the amount of a compute resource and the resource-to-resource ratio. For example, Compute intensive instances may have a higher CPU to random-access memory (RAM) ratio, while memory intensive instances may have higher RAM to CPU ratio, and similarly with all other resources.

The choice of resource amount and ratio is of vital importance, as it has a crucial impact on the performance of the machine for a workload configuration\cite{jackson2010performance}. Given the number of possible workload configurations and the ever-growing list of instance types, it has become paramount in today's cloud computing environment to design a method for evaluating workload/machine combinations. Alipourfard \textit{et al.}~\cite{alipourfard2017cherrypick} address this issue by introducing CherryPick, a system that uses Bayesian Optimization to build performance models for various workloads that distinguish optimal or close to optimal VMs from the rest, with only a few test runs per workload configuration.
	
The different types of workload configurations can have varying effects on the physical computing resources seen through the resource utilization data. Understanding how different workload configurations affect computing resources is crucial for proper resource allocation. To efficiently allocate system resources for a workload, the capability to properly predict the characterization of a workload on the computing resource is essential whether that be for on-premises or on cloud computing environment. Koh \textit{et al.}~\cite{koh2007analysis} use workload characterization by studying the effects of performance at system-level workload characteristics. By analyzing the collected data, they were able to identify different application clusters that generate certain types of performance. They subsequently developed models to predict the performance of a new application from its workload characterization. Khan \textit{et al.} \cite{khan2012workload} identify repeatable workload patterns by exploring cross-VM workload correlations resulting from the dependencies among applications running on different VMs. By treating workload data samples as time series, they used a clustering technique to identify groups of VMs that exhibit correlated workload patterns. Then, they used a method based on hidden Markov models (HMM) to characterize the temporal correlations in the discovered VM clusters and to predict variations of workload patterns.
	
Understanding the characterizations of a workload configuration and how they affect computing resources is essential for improving compute resource allocation for the workload. However, customers of cloud computing providers are tasked with choosing a VM instance type from a large list of combinations of family types and sizes. Not only is the choice prohibitively difficult, but has enormous implications on both performance and cost for the given workload configuration. A fundamental void exists for a performance indicator that accurately, economically and consistently provides a performance score for a given workload and computing machine configuration. Such a performance indicator would help determine the computing configuration that would perform best with the workload configuration. Hsu \textit{et al.}~\cite{hsu2018micky} show that there is often a single cloud configuration that is surprisingly near-optimal for most workloads. They introduce a collective-optimizer, MICKY, that reformulates the task of finding the near-optimal cloud configuration as a multi-armed bandit problem. MICKY efficiently balances exploration of new cloud configurations and exploitation of known good cloud configuration. More relevant work on the challenge of mapping workload categorizations to physical resources can be found in~\cite{pietri2016mapping}. Yadwadkar \textit{et al.}~\cite{yadwadkar2017selecting} address the problem of optimal VM selection with PARIS, a data-driven system that uses a hybrid offline and online data collection and modeling framework. PARIS predicts workload performance for different user-specified metrics, and also the resulting costs for a wide range of VM types and workloads across multiple cloud providers. While the aforementioned methods deliver a VM, the proposed WISE score, however, tells us whether we need to change our configuration or if it is good choice.
	
The proliferation of cloud computing availability and the mass adoption of cloud computing as a viable option to on-premises computing have triggered the need for workload/machine performance indicators. Cloud computing offers high availability, scalable, efficient and cost saving computing performance for any workload configuration. These considerations require resource planning to determine the required compute resources and how/when compute resources need to grow or be scaled back. Compute resource planning requires a good understanding of both the computing resources capacities and the workloads resource utilization patterns. Rjoub \textit{et al.}~\cite{rjoub2020bigtrustscheduling} address the task of guaranteeing performance while minimizing resource utilization from a task scheduling perspective. They propose a trust-aware scheduling solution, which consists of VMs' trust level computation, tasks priority level determination, and trust-aware scheduling. Also, Rjoub \textit{et al.}~\cite{rjoub2019deep} present an automated big data task scheduling approach for cloud computing environments. The approaches introduced in~\cite{rjoub2020bigtrustscheduling,rjoub2019deep} are for optimum resource utilization from a task scheduling perspective, whereas in our work we describe methods for scoring the performance of a workload on a particular machine.

From a workload characterization perspective, Mishra \textit{et al.}~\cite{mishra2010towards} present an approach for workload classification by identifying the workload dimensions, constructing task classes using a clustering algorithm, determining the break points for qualitative coordinates within the workload dimensions, and merging adjacent task classes to reduce the number of workloads. They show that the duration of a task is bimodal, has either a short or a long duration, and that most tasks have a short duration. Also, that most compute resources are consumed by a few tasks with a long duration that have large demands for CPU and memory capacity. Downey \textit{et al.}~\cite{downey1999elusive} present a characterization of a workload on a parallel system from an operating system perspective by investigating means to characterize the stream of parallel jobs submitted to the system, their resource requirements, and their behavior. A comprehensive survey of workload characterization can be found in~\cite{calzarossa1993workload}.
	
In this paper, we present an integrated approach to derive computing machine and computing resource performance indicators for a given workload configuration. The proposed WISE framework indicates how well a computing machine is running with a specific workload configuration, and whether a different computing configuration could get better performance. Given the need for a computing performance indicator for a given workload configuration, we describe a novel method  for scoring the performance of the machine while running the workload, meaning is the machine being under-utilized, over-utilized or is it running in a sweet spot? In particular, we consider how far off from described target levels the resources are running at, i.e. whether there exists resource waste or strain.

The rest of this paper is organized as follows. In Section 2, we present the WISE framework for scoring the performance of a machine, given a specific workload, and we describe its main algorithmic steps. Experimental results using the WISE framework on distinct workloads are presented in Section 3. Finally, we conclude in Section 4 and point out some future work directions.
	
\section{Method}
\begin{figure}
\includegraphics[scale=0.35]{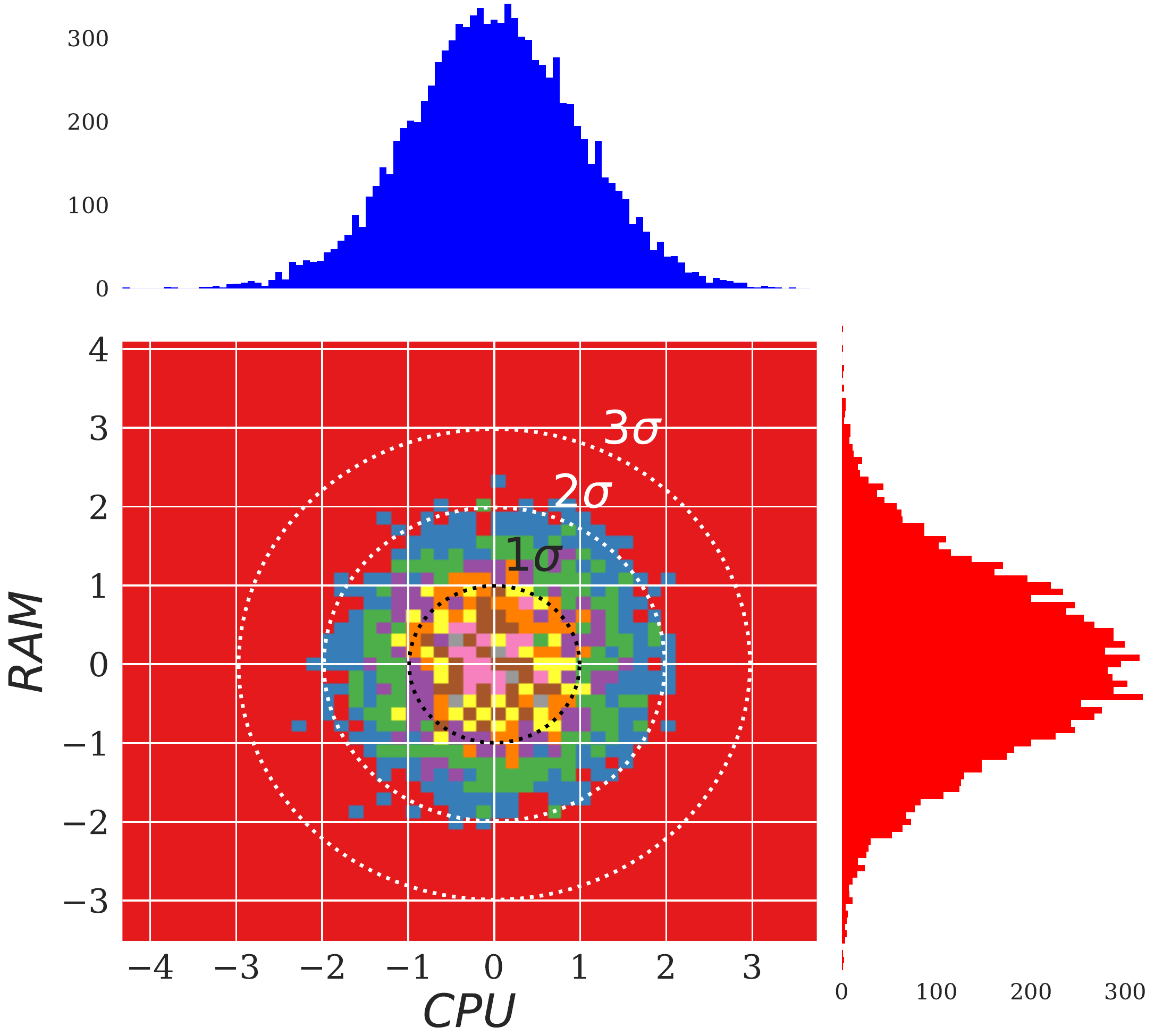}
\caption{Illustration of machine scoring in two dimensions (CPU and memory), with machines closer to the targets having a value closer to 1.}
\label{Fig:wmi_score_plot}
\end{figure}

In this section, we introduce the WISE framework for computing a fitness score for a workload/machine combination. The proposed approach encompasses described resource target levels and ranges, as well as the utilization distances from these targets into a single metric. This performance metric is used to easily determine how well a workload/machine combination is performing and also to diagnose possible issues. A key advantage of WISE is that we can use as many computing resources as necessary. In addition, computing resource can use as many aggregate utilization percentage rates as desired. Figure~\ref{Fig:wmi_score_plot} illustrates hypothetical data
points in two dimensions (CPU and memory), with machines closer to the targets having a value closer to 1. The WISE score encompasses this information into an index score that indicates the performance of the machine given a workload.

For each computing resource aggregate combination, we first define a target level and range, and then set the limits as to what are acceptable running rates for such a  resource using that aggregation. For example, in the case of CPU we can set the cpu utilization average running target at 40\% with a range between -30\% and 30\%, meaning that any rate between 10\% and 70\% utilization is deemed to be acceptable, with a rate closer to 40\% being better.

The benefits of the proposed machine scoring approach may be summarized as follows:
\begin{itemize}
\item Any number of resources can be used (CPU, memory, network, disk, etc..) depending on specific need or for general machine usage and any number of aggregations can be used for each resource (average, p95, p50, etc..). More aggregates allow a more specific machine-workload configuration. We discuss this in further detail in the \textit{Discussions} section.

\item Optimal machines are defined as ones falling within the defined acceptable range, but any range can be used depending on need.
\item The WISE framework can be used to train models on machines that have scores above a certain threshold for better performance, as the models would be trained on machines that are running between defined ideal levels.
\item Workload Validation is more thorough, as the WISE score allows evaluation of machines where utilization data exists from running a specific workload. This is demonstrated in the \textit{Validation} section.
\item Users can define their own ideal running machine/resource rates and ranges, and subsequently evaluate their machines using the WISE score.
\item Getting a WISE score pre and post configuration (vm or machine) change will validate the benefits with clear indicators.
\item The individual resources' scores give an indication of how a resource is running with respect to the defined targets and ranges. For instance, if a machine has a low score, then we can look at the individual resources' scores and locate the origin of the problem.
\end{itemize}
	
\medskip\noindent{\textbf{Global vs Machine Specific:}}\quad The running levels described next can be global or machine specific. Global resource running levels are applied to all machines, while machine specific types are applied only the that specific instance type. The machine specific defined running levels override the global levels for that machine specific type.

\subsection{Setting Targets and Ranges}
How to set good target and ranges for the different computing resources is one of the challenging issues to tackle while looking at computing performance given different workloads. The answer to what are acceptable performance levels will depend heavily on the use case\cite{lilja2005measuring,henning2000spec}. The targets and ranges are set after careful consideration on what the performance expectations are for the given a workload and for what the tolerance level of computing waste and/or unexpected workload changes are. Varying use cases may have different targets and ranges depending on tolerance. For instance, non-critical workloads have higher targets and ranges, as a slowdown due to an unexpected spike is not damaging (e.g. an online library or non-mission critical systems). In addition, critical workloads add more leeway in their targets and ranges to account for unexpected spikes in order to make sure there is no degradation in workload performance (e.g. in real time systems or mission critical systems).

\subsection{Algorithmic Steps}
In this subsection, we describe in detail the main algorithmic steps of the proposed WISE framework.
\subsubsection{Ideal Resource Running Levels}
In this first step, we define the ideal running levels for each resource $r_i$, where $i=1,\dots,n$ and $n$ is the total number of resources. More precisely, for each resource $r_i$ (e.g. average memory utilization), we define the following parameters:
\begin{itemize}
\item The ideal target running utilization level $\mu_i$ for the given resource ($r_i$).
\item The acceptable deviation levels $\sigma_i$ from the ideal target level $\mu_i$ for the given resource $r_i$.
\item An upper limit $r^{\max}_i$ for the given resource $r_i$. Resources running above this level will be penalized with a zero score.
\item If there is no ideal target running utilization level $\mu_i$ for the given resource $r_i$, but there is an upper limit $r^{\max}_i$, set the target and range to $0$. Resources running above this level will be penalized with a zero score, otherwise the resource will have no affect on the score.
\end{itemize}
As an example, if we consider the average memory utilization as a resource with parameters $\mu = 50\%$, $\sigma = 30\%$ and $r^{\max}_i = 90\%$, then the average memory utilization running between $20\%$ and $80\%$ would be acceptable, with levels running closer to the defined target having a higher score. The further the running level is from the target level, the worse the score will be for that resource. Machines having a running level above the $r^{\max}$ (e.g. $90\%$ or above) would have the worst possible score for that resource.
	
\begin{figure}[!htb]
\setlength{\tabcolsep}{.1em}
\centering
\begin{tabular}{c}
\includegraphics[scale=0.65]{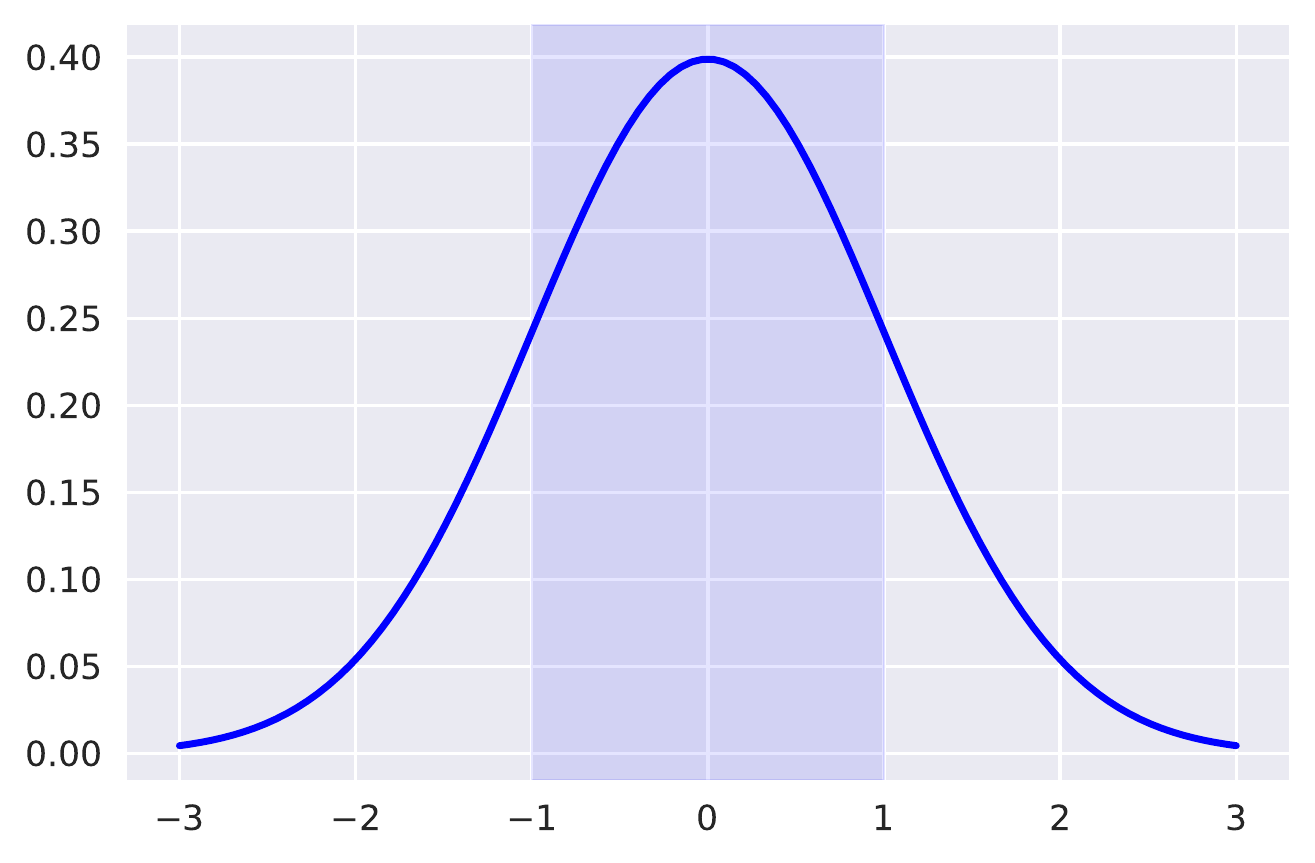} \\
\includegraphics[scale=0.65]{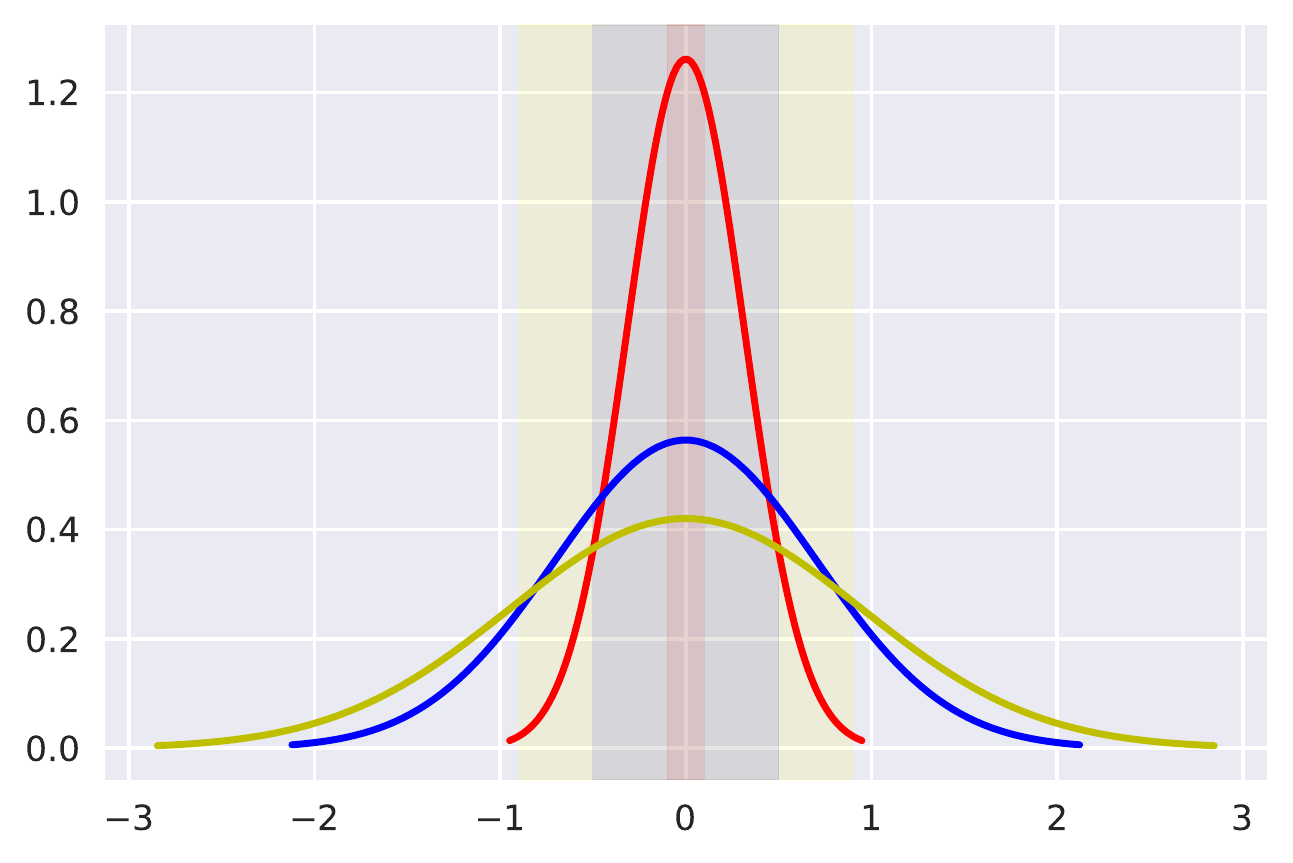}
\end{tabular}
\caption{Top: Defining a target and standard deviation. Bottom: Plot with different targets and standard deviations.}
\label{fig:targets}
\end{figure}

Figure~\ref{fig:targets} (top) displays a normal distribution with target $\mu=0$ and standard deviation $\sigma$ equal to $1$. It shows how a target level and ranges can be defined for resources. The darker blue area indicates ideal running levels, while the lighter blue indicates substandard running levels. Figure~\ref{fig:targets} (bottom) shows a plot of normal distributions with varying target levels and standard deviations.
	
\subsubsection{Deviations from Ideal Running Levels}
In this second step, we define the resource z-score (also known as standard score) as
\begin{equation}
z_i=\frac{x_i-\mu_i}{\sigma_i},\quad i=1,\dots,n,
\end{equation}
where $x_i$ is the resource utilization rate, which is the actual utilization rate for a given resource $r_i$. The resource z-score gives the number of standard deviations from the target $\mu_i$ for each resource $r_i$.
	
Continuing with the example above, if a machine has an an average memory utilization running level of $80\%$, then its z-score is equal to 1, while a machine with a running level of $20\%$ has a z-score of -1.
	
\subsubsection{Resources Scores}
In this third step, we define a resource score $s_i$ for each resource $r_i$ using the hyperbolic tangent and the exponentially monotone functions. The resource score provides a normalized index value for each resource. The resource score gives an index value for each resource, where the utilization rate falls within a normalized range. The closer the utilization is to the target the better the score will be and the further it is away from the target the worse the score will be.

\begin{figure}[!htb]
\setlength{\tabcolsep}{.1em}
\centering
\begin{tabular}{c}
\includegraphics[scale=0.57]{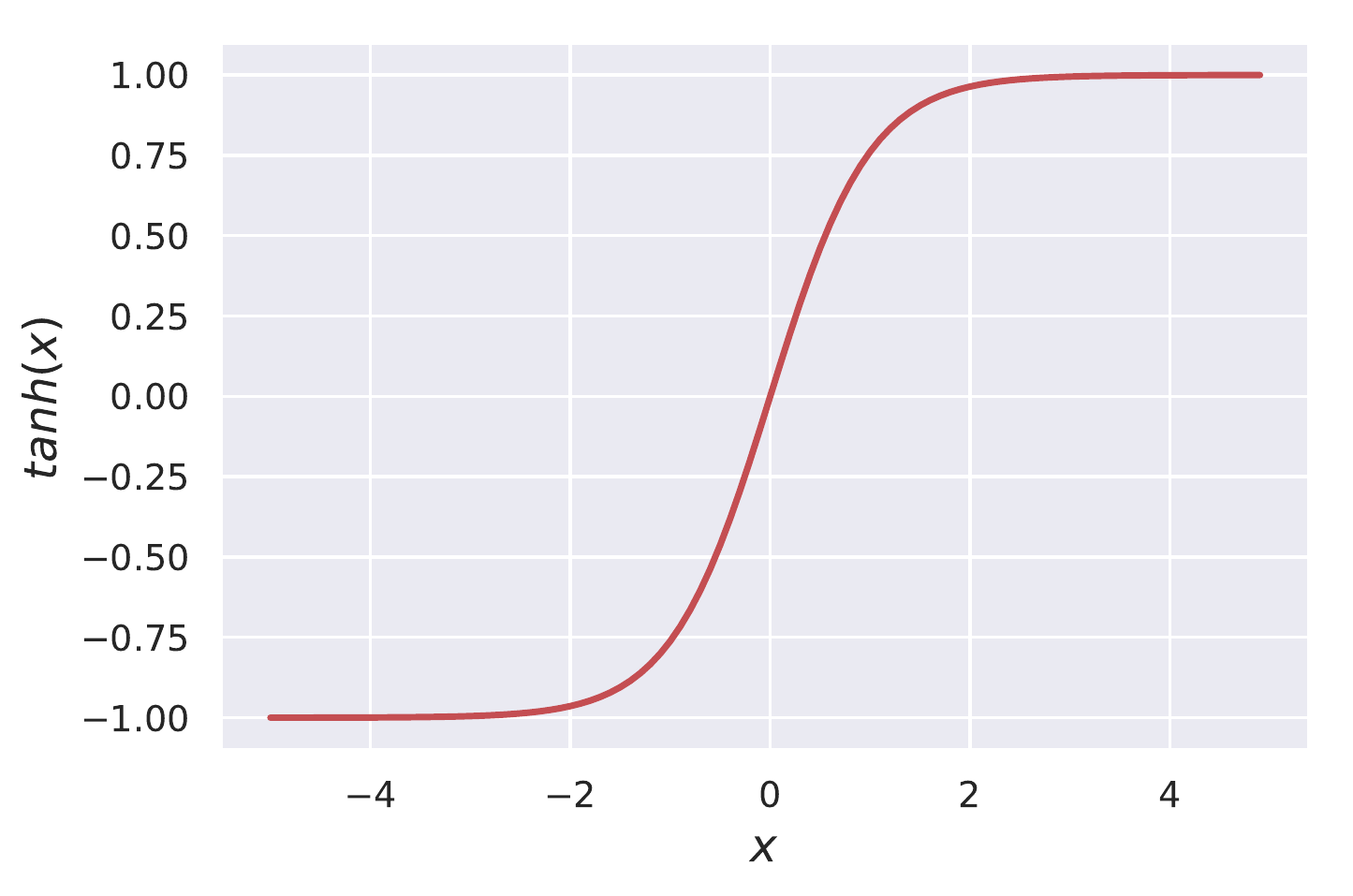} \\
\includegraphics[scale=0.55]{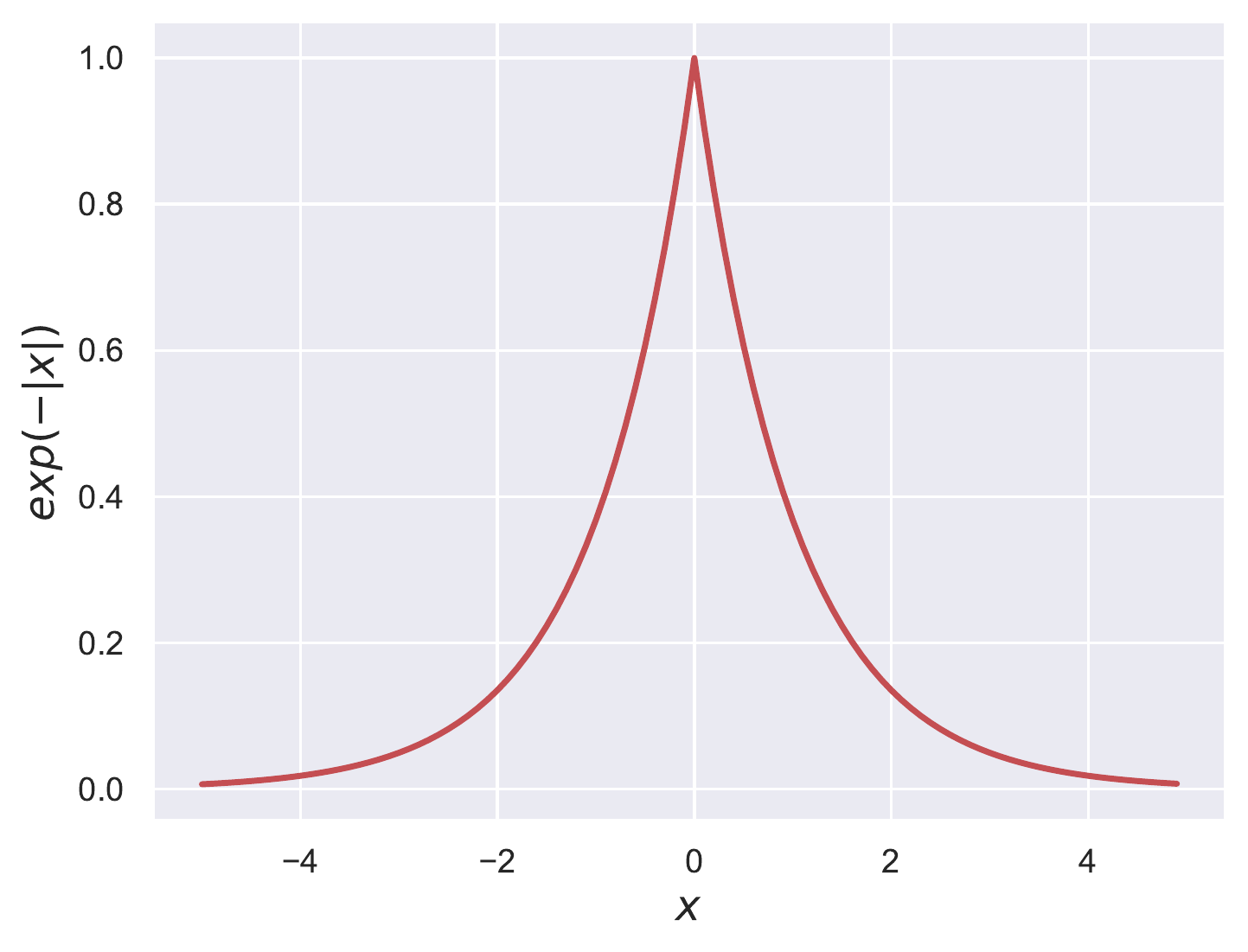}
\end{tabular}
\caption{Top: Hyperbolic tangent function. Bottom: Exponentially monotone function.}
\label{Fig:functions}
\end{figure}

\medskip\noindent{\textbf{Scoring functions:}}\quad The scoring functions serve two main purposes: (1) Standardize the score to within a given level for both resources and the overall score. A scoring function converts the resource z-score to a predetermined range for all resources. This allows users of the WISE score for both resources and the overall score to examine the score without knowing a priori what the targets and ranges are; (2) Limit the negative influence that a sub-optimal performing resource can have on the overall score. The scoring functions limit the adverse effect that one resource can have on the overall score by smoothing the score when the resource z-score gets larger, as illustrated in Figure \ref{Fig:functions}. Such a situation may arise when a resource is running way off the target and range levels, resulting in a high z-score value. This would cause that specific resource and the overall machine to have sub-optimal WISE scores without a scoring function, when in fact if all other resources are running perfectly that one resource should not determine the overall score outcome. To circumvent this limitation, we introduced a \textit{penalty term} in the WISE framework to deal with the situation where a resource running above a certain level should cause the overall WISE score to be at a sub-optimal level to indicate potential issues that need to be addressed.

\medskip\noindent{\textbf{Using the hyperbolic tangent function:}}\quad Given a resource z-score, we define the resource score as
\begin{equation}
s^{\tanh}_i = \tanh(z_i), \quad i=1,\dots,n,
\end{equation}
where $\tanh(\cdot)$ is the hyperbolic tangent function, which is commonly used as an activation function in neural networks and it produces outputs in the scale of $[-1, 1]$, as shown in Figure~\ref{Fig:functions} (top). A negative value indicates a resource or machine utilization rate below the ideal rate, while a positive value indicates a utilization rate above the target rate. A value of 0 indicates that the resource is running at the target level. A value of -1 indicates a resource that is at the extreme of under-utilization and a value of +1 indicates a resource at the extreme of over-utilization. Given this range of values for both the computing resources and an overall value for the machine allows customers to diagnose possible issues with the machine and/or resource usage.

The interesting aspect of using the hyperbolic tangent function to compute the resource score is that a negative score indicates an under-utilized resource with respect to the ideal target level, while a positive score indicates an over-utilized resource with respect to the ideal target level. This indication for each resource can be very helpful in diagnosing various issues with the computing machine and/or resources.

\medskip\noindent{\textbf{Using the exponentially monotone function:}}\quad For each resource z-score, we define the resource score as
\begin{equation}
s^{\exp}_i = \exp(-|z_i|), \quad i=1,\dots,n,
\end{equation}
where $\exp(-|t|)$ is the exponentially monotone function shown in Figure~\ref{Fig:functions} (bottom). This function produces outputs in the scale of $[0, 1]$, where the best score is 1 and the worst is 0. The interesting aspect of this score is that the best value would be 1 or close to one and the worst is 0, which is quite intuitive.

Notice that using the absolute value of $z_i$, the output of the exponentially monotone function is between 0 and 1. A value of 1 indicates that the resource is running at the target level. A value of 0 indicates a resource that is at the extreme of under- or over-utilization.

\medskip\noindent{\textbf{Resource weight:}}\quad It is important to note that not all resources have equal importance on the computing environment. For instance, CPU and memory might have more influence on how well or bad a machine is behaving. This can also vary depending on the workload demands and goals on the machine. To this end, a predetermined weight $w_i$ is assigned to a given resource score $s_i$. This weight can be used if we notice that certain computing resources, such as memory and CPU, are more important to the health of a machine than for example networking or disk. When no weight is assigned, all resources have equal weight in the resource score.

\subsubsection{Resource Penalty}
In this fourth step, we introduce a resource penalty term, which affects the machine score when a given resource surpasses a maximum threshold. The reason behind using a penalty term is largely due to the fact that when some resources are stretched to a certain maximum level, the whole machine suffers and becomes unusable. We want to penalize resources running above an upper limit $r^{\max}_i$ for the given resource so that the machine score is negatively influenced. To this end, we first subtract the upper limit $r^{\max}_i$ from the resource running level $x_i$ to get a positive value if the resource is running above the resource upper limit, 0 if it is equal to it and negative if it is running below it, i.e.
\begin{equation}
\sgn(x_i-r^{\max}_i) = \left\{
\begin{array}{lll}
-1 & \quad x_i < r^{\max}_i \\
0 & \quad x_i = r^{\max}_i \\
1 & \quad x_i > r^{\max}_i
\end{array}
\right.
\end{equation}
where $\sgn(\cdot)$ is the sign function.

\medskip\noindent We define the penalty term for each resource as
\begin{equation}
\mathcal{P}(x_i)=H(x_i-r^{\max}_i),\quad i=1,\dots,n,
\end{equation}
where
\begin{equation}
H(t) = \left\{
\begin{array}{ll}
1 & \quad t \geq 0 \\
0 & \quad t < 0
\end{array}
\right.
\end{equation}
is the Heaviside function (also referred to as unit step function). It is worth pointing out that the sign and Heaviside functions are related via the identity: $\sgn(t)=H(t)-H(-t)$.

The penalty term returns a value of 0 or 1 depending on whether that resource is running above or below the pre-defined max utilization rate $r^{\max}_i$. If the resource in not above the resource limit, then the penalty term has a value of 0, and hence it does not have any affect on the resource score. On the other hand, if the resource utilization is above the resource limit, then the penalty term for that resource is equal to 1.

Using nonnegative penalty weight factor $\alpha$ for each resource $r_i$, we define the weighted penalty term for each resource as
\begin{equation}
\mathcal{P}_{\alpha}(x_i)=\alpha H(x_i-r^{\max}_i),\quad i=1,\dots,n,
\end{equation}		
where an $\alpha$ value of 1 sets the resource penalty term to 1 for values greater than or equal to 1 and 0 otherwise. An $\alpha$ value smaller than 1 diminishes the affect of the penalty term, while an $\alpha$ value larger than $1$ increases the affect of the penalty term. A high value for $\alpha$ ensures that a machine with a resource that is running at levels over the $r^{\max}_i$ receives the worst possible machine score.	
	
\subsubsection{Workload/Machine Index Score}
In this last step, we introduce four variants of the proposed WISE score using the hyperbolic tangent and exponentially monotone functions in conjunction with the weighted $\ell_1$- and $\ell_2$-norms. The WISE score gives a good indication on how well the computing machine is running given a specific workload.

\medskip\noindent{\textbf{Using tanh function and $\ell_1$-norm:}}\quad We define the WISE score as
\begin{equation}
\mathcal{S}_1= \min\left[\frac{1}{n}\sum_{i=1}^{n}w_i \left|\tanh(z_i)\right|+ \sum_{i=1}^{n} \mathcal{P}_{\alpha}(x_i),1\right],
\end{equation}	
where the minimum function is used to assure that a value greater than 1, which is the worst score on the positive side, is not returned even when more than one resource utilization rate falls above its upper limit rate.

The penalty term adds a value depending on whether or not any of the resource utilization rates are above their respective upper limit rate. If none of the resources is above its upper limit rate, then the penalty term is $0$ and hence it has no affect on the WISE score. On the other hand, if there is at least one or more resources that are above their respective upper limit rates, then the penalty term has a value greater than 1, which adversely affects the machine score, indicating over-utilization. The WISE score has a value between 0 and 1, with a value of 0 being the best.
	
\medskip\noindent{\textbf{Using tanh function and $\ell_2$-norm:}}\quad We define the WISE score as
\begin{equation}
\mathcal{S}_2 = \min \left[\frac{1}{n}\sqrt{\sum_{i=1}^{n}\big(w_i \tanh(z_i)\big)^2} + \sum_{i=1}^{n} \mathcal{P}_{\alpha}(x_i), 1 \right],
\end{equation}
which returns values between 0 and 1, with a value of 0 being the best.
	
\medskip\noindent{\textbf{Using exponentially monotone function and $\ell_1$-norm:}}\quad We define the WISE score as
\begin{equation}
\mathcal{S}_3= \max\left[\frac{1}{n}\sum_{i=1}^{n}w_i e^{-|z_i|}- \sum_{i=1}^{n} \mathcal{P}_{\alpha}(x_i),0\right],
\end{equation}	
where the maximum function is used to assure that the machine score is nonnegative, with a value of 0 being the worst score. The exponentially monotone function returns a value between 0 and 1 with 1 being the best), while the penalty term returns a value between 0 and the number of resources $n$ times the parameter $\alpha$, depending on the number of resource utilization rates that fall above the upper limit rate.
	
\medskip\noindent{\textbf{Using exponentially monotone function and $\ell_2$-norm:}}\quad We define the WISE score as
\begin{equation}
\mathcal{S}_4= \max\left[\frac{1}{n}\sqrt{\sum_{i=1}^{n}\bigl(w_i e^{-|z_i|}\bigr)^2}- \sum_{i=1}^{n} \mathcal{P}_{\alpha}(x_i),0\right],
\end{equation}	
	
\section{Experiments}
In this section, we conduct extensive experiments by running a two distinct workloads on multiple Amazon AWS EC2\footnote{https://aws.amazon.com/ec2} instances using the utilization data generated to compare the WISE scores for each instance. We evaluate the WISE score on two benchmarks, a MongoDB workload which is a cpu intensive workload and a second Streaming workload which is a networking intensive workload. To account for bursty workloads in the \textit{Validation} section, we add a new third workload which is a reconfiguration of the MongoDB workload to have bursts of activity with periods of less activity.

\subsection{Experimental Setup}
In all the experiments, we set the penalty weight factor $\alpha$ to 1 and used a uniform resource weight. We considered five resources ($r$) with target resource utilization levels ($\mu$), acceptable deviation levels ($\sigma$) and upper limits ($r^{\max}$) as described in Table \ref{Tab:Experiment_Parameters}, which describes all the parameters that are used for the WISE score calculation. The optimal thresholds are set to the levels for one range ($\sigma$) deviation from the target ($\mu$) as described in Table \ref{Tab:Experiment_Thresholds},  although these can vary depending on a users use case. The table describes thresholds used for both functions (TanH and Exp.), which are used to determine if the workload/machine is performing within an acceptable range.

\begin{table*}[!htb]
	\rowcolors{1}{}{lightblue}
	\centering
	\caption{Parameters used in the experiments and validation.}
	\begin{tabular}{l*{6}{c}}
		\toprule
		Resource ($r$) & Target ($\mu$) & Range ($\sigma$) & Weight ($w$) & Resource Max ($r^{\max}$) & Penalty Weight ($\alpha$)\\
		\midrule
		CPU/Avg & 40\% & 30\% & 1 & N/A & 1 \\
		CPU/P95 & 70\% & 20\% & 1 & N/A & 1 \\
		RAM/Avg & 50\% & 20\% & 1 & 90\%  & 1  \\
		RAM/P95 & 70\% & 30\% & 1 & N/A  & 1 \\
		Network/Avg & N/A & N/A & 1 & 80\%  & 1 \\
          \bottomrule
	\end{tabular}
	\label{Tab:Experiment_Parameters}
\end{table*}

\begin{table}[!htb]
	\rowcolors{1}{}{lightblue}
	\centering
	\medskip
	\caption{Optimal thresholds used in the experiments and validation.}
	\begin{tabular}{l*{3}{c}}
		\toprule
		Function & WISE Score & Threshold & Best Score\\
		\midrule
		TanH   & Resource & $-0.76 \leq r \leq 0.76$ & 0\\
		TanH   & Overall & $r \leq 0.76$ & 0\\
		Exp.   & Resource & $r \geq 0.36$ & 1\\
		Exp.   & Overall & $r \geq 0.36$ & 1\\
        \bottomrule
	\end{tabular}
	\label{Tab:Experiment_Thresholds}
\end{table}

\subsection{Results}
\begin{figure*}[!htb]
\centering
\includegraphics[scale=0.35]{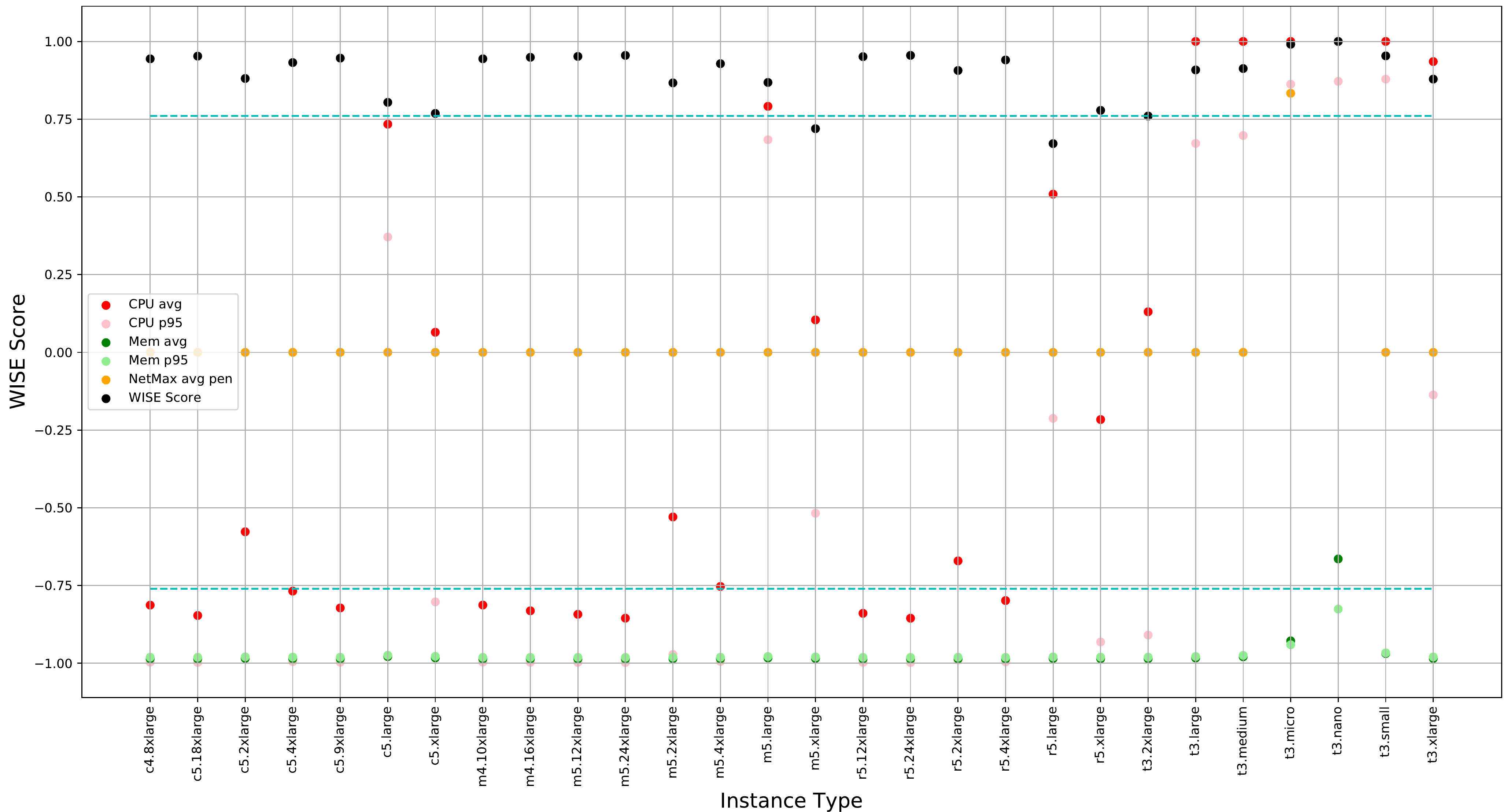}
\caption{WISE scores using the hyperbolic tangent function and $\ell_1$-norm for MongoDB Workload.}
\label{fig:wl1_tanh_avg}
\end{figure*}

\begin{figure*}[!htb]
\centering
\includegraphics[scale=0.35]{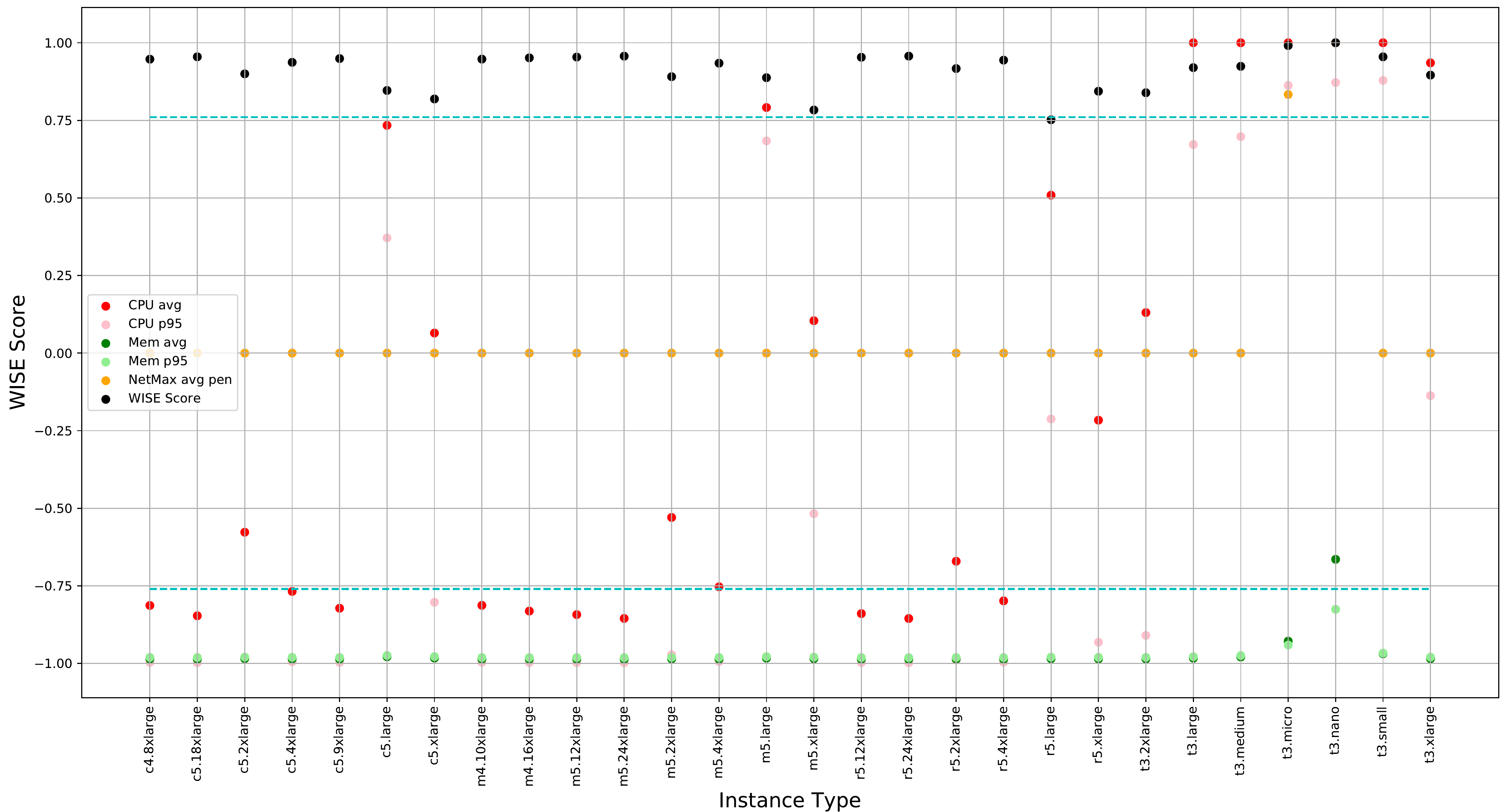}
\caption{WISE scores using the hyperbolic tangent function and $\ell_2$-norm for MongoDB Workload.}
\label{fig:wl1_tanh_eucl}
\end{figure*}

In this subsection, we demonstrate the performance of our proposed WISE framework on two distinct workload configurations. Figure~\ref{fig:wl1_tanh_avg} displays the WISE score and resource scores for a MongoDB workload using the hyperbolic tangent function and $\ell_1$-norm, with the best score having a value of 0. The area between the light blue dotted lines indicate values that fall within the described acceptable ranges. Any score that falls outside of this region indicates over- or under-utilization from the acceptable ranges. A machine score can still fall within the acceptable range while having a resource that falls out of range. As average network utilization has only an upper limit (penalty), it can only affect the machine score when the utilization surpasses this limit. This is displayed in Figure~\ref{fig:wl1_tanh_avg} by showing a dot on 0 when it has no affect. Figure \ref{fig:wl1_tanh_eucl} displays the same data as in Figure \ref{fig:wl1_tanh_avg}, except that the overall machine score is computed using the hyperbolic tangent function and $\ell_2$-norm.


\begin{figure*}[!htb]
\centering
\includegraphics[scale=0.35]{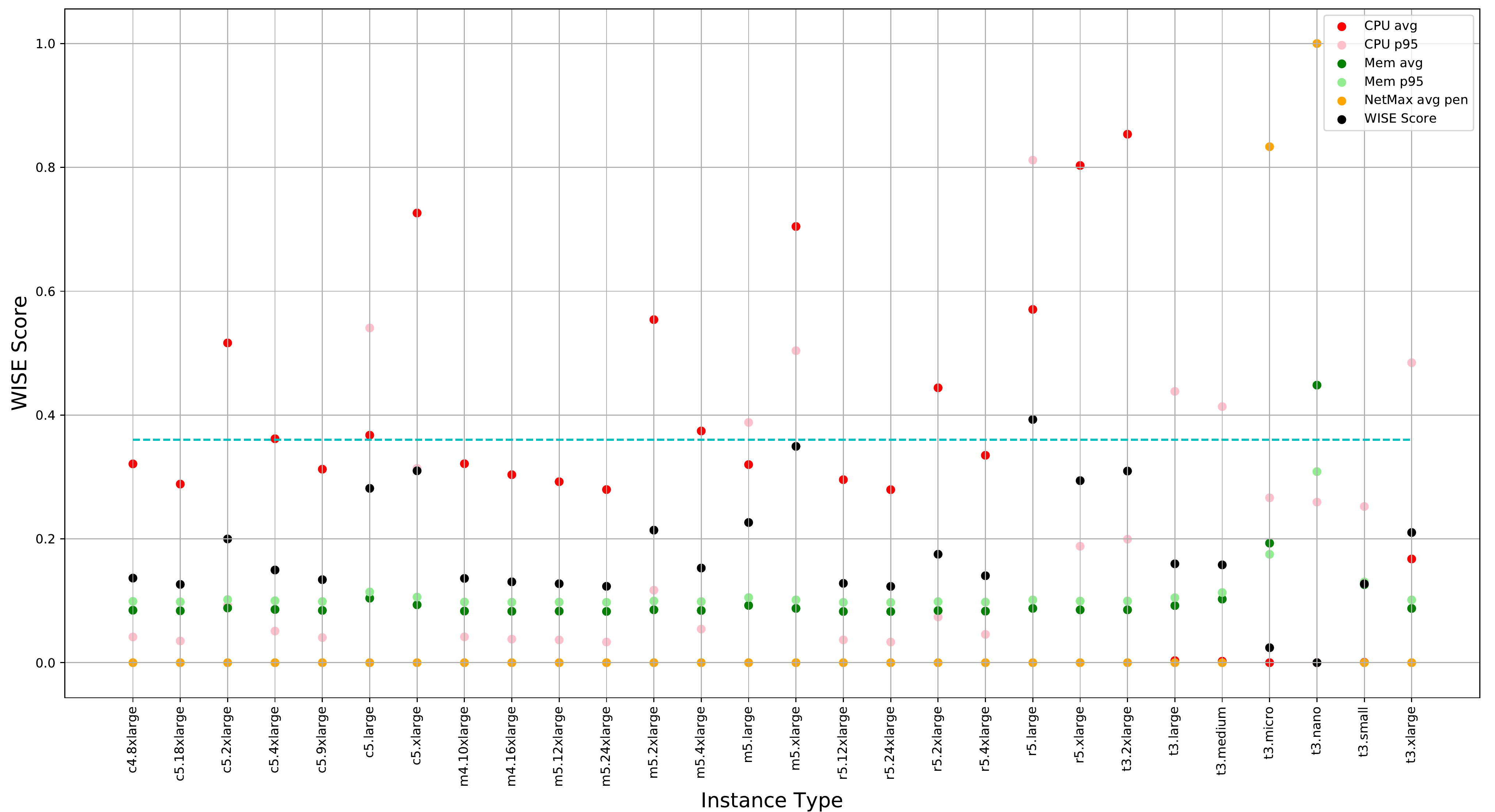}
\caption{WISE scores using the exponentially monotone function and $\ell_1$-norm for MongoDB Workload.}
\label{fig:wl1_exp_avg}
\end{figure*}

\begin{figure*}[!htb]
\centering
\includegraphics[scale=0.35]{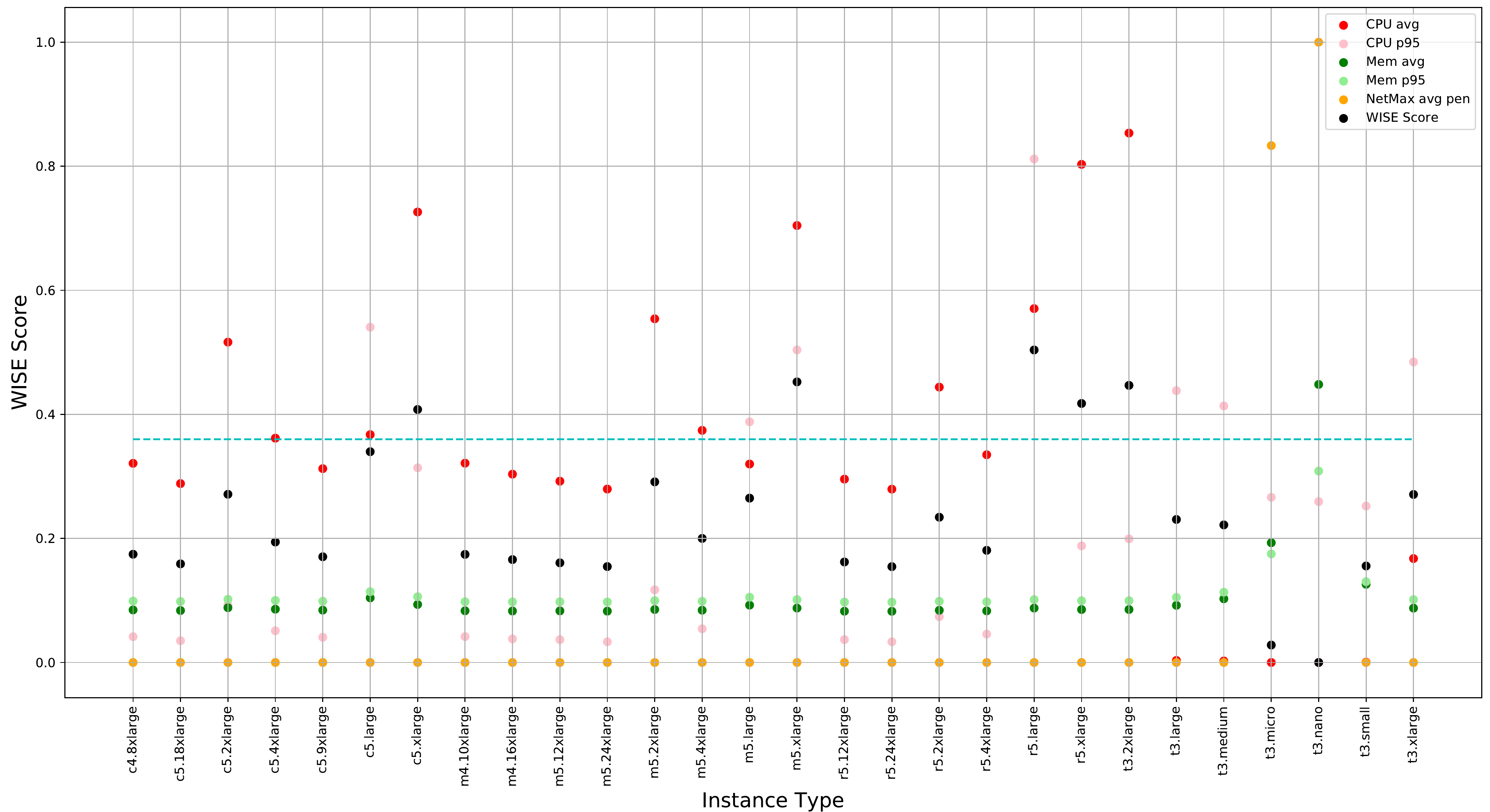}
\caption{WISE scores using the exponentially monotone function and $\ell_2$-norm for MongoDB Workload.}
\label{fig:wl1_exp_eucl}
\end{figure*}

Figure~\ref{fig:wl1_exp_avg} also displays the WISE score and resource scores for a MongoDB workload using the exponentially monotone function and $\ell_1$-norm, with the best score having a value of 1. The area above the light blue dotted lines indicate values that fall within the described acceptable ranges. Any value that falls below this region indicates over- or under-utilization from the acceptable ranges. A machine score can still fall within the acceptable range while having a resource that falls out of range. Figure~\ref{fig:wl1_exp_eucl} displays the same data as in Figure~\ref{fig:wl1_exp_avg}, except that the overall machine score is computed using the exponentially monotone function and $\ell_2$-norm.

\begin{figure*}[!htb]
\centering
\includegraphics[scale=0.35]{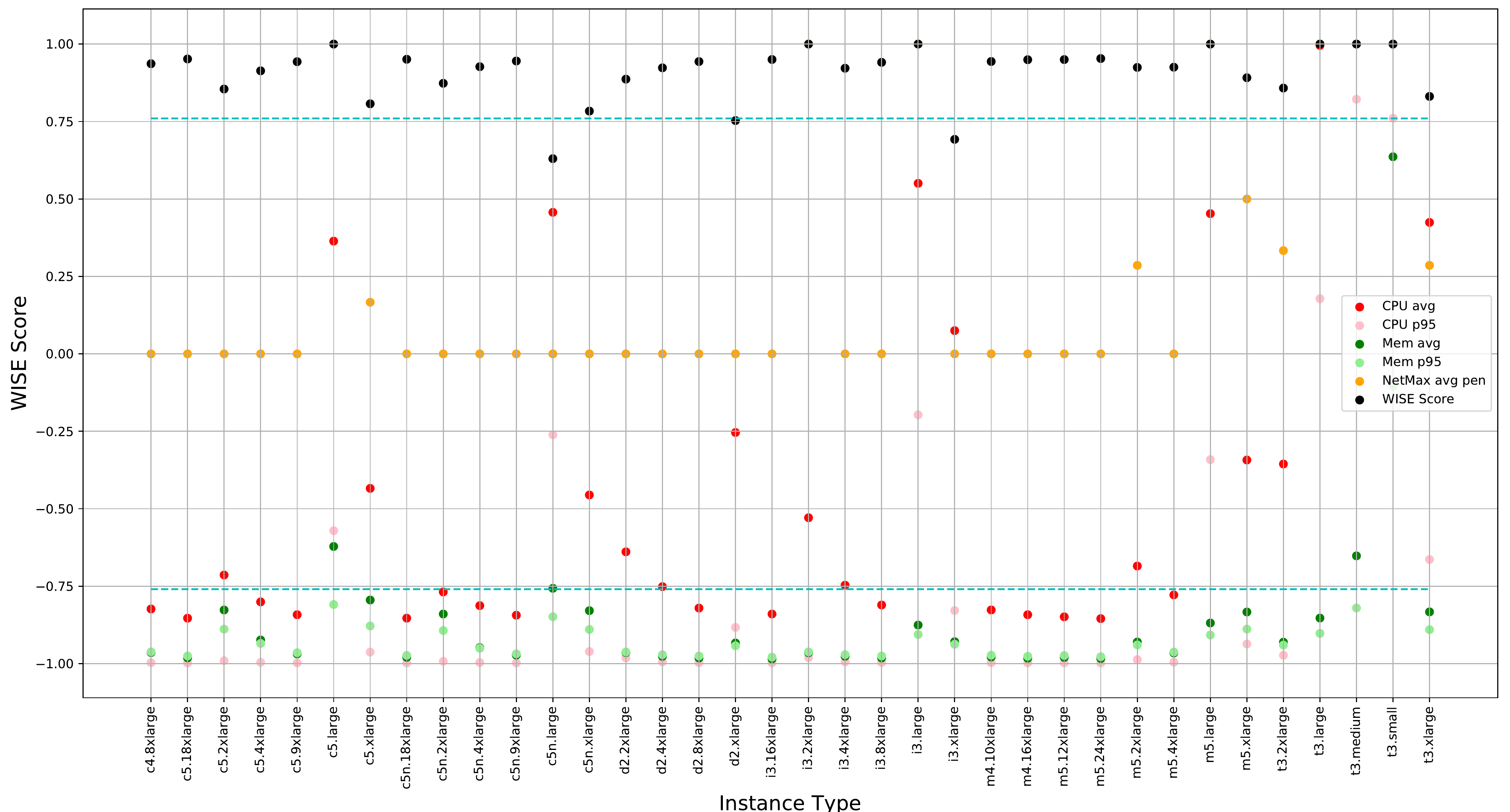}
\caption{WISE scores using the hyperbolic tangent function and $\ell_1$-norm for Streaming Workload.}
\label{fig:wl2_tanh_avg}
\end{figure*}

\begin{figure*}[!htb]
\centering
\includegraphics[scale=0.35]{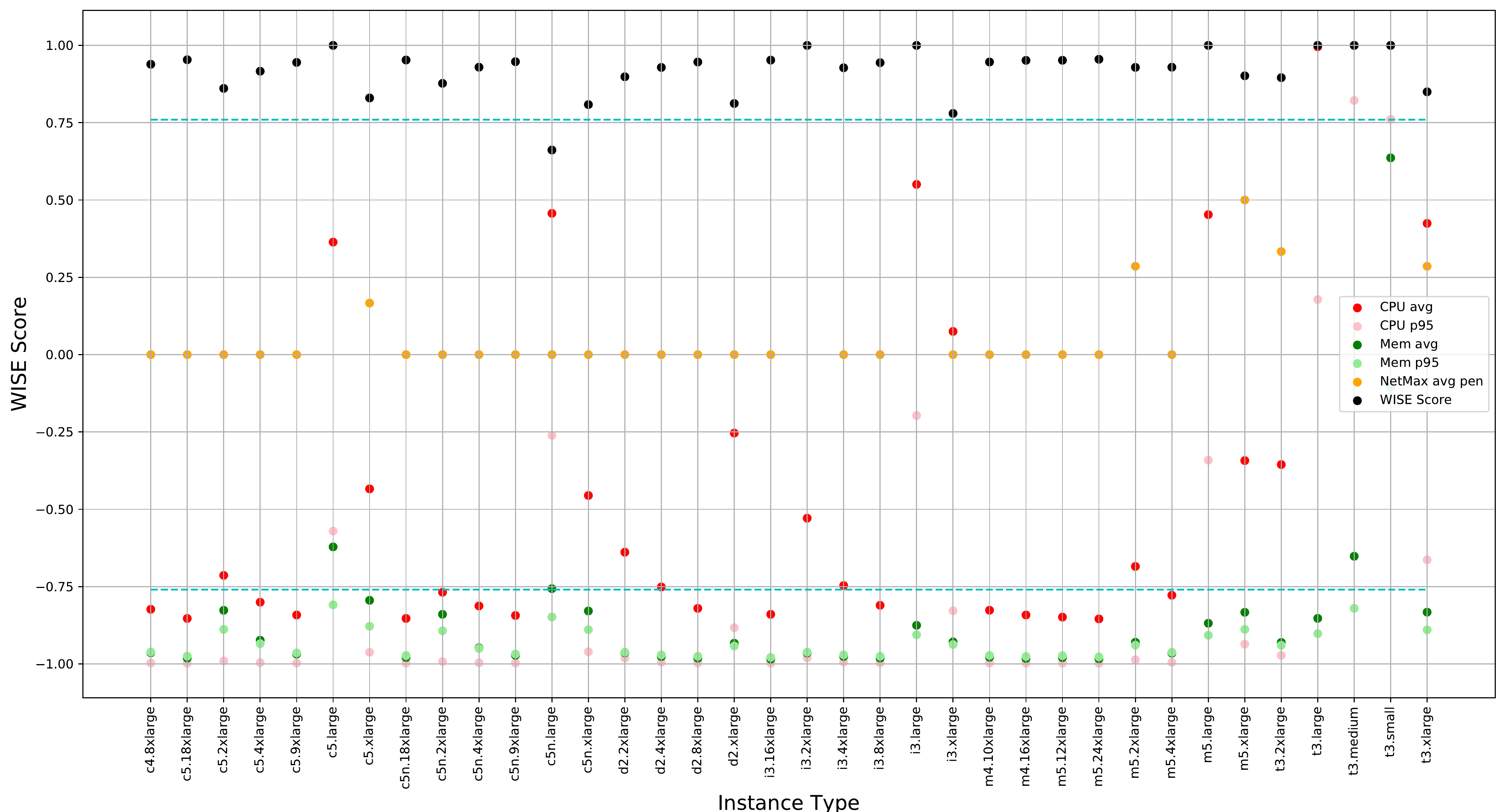}
\caption{WISE scores using the hyperbolic tangent function and $\ell_2$-norm for Streaming Workload.}
\label{fig:wl2_tanh_eucl}
\end{figure*}

Figure \ref{fig:wl2_tanh_avg} displays the WISE score and resource scores for a Streaming workload. The normalization function using the hyperbolic tangent function and $\ell_2$-norm, with the best score having a value of 0. The area between the light blue dotted lines indicates values that fall within the described acceptable ranges. Any value that falls outside of this region indicates over- or under-utilization from the acceptable ranges. A machine score can still fall within the acceptable range while having a resource that falls out of range. As average network utilization has only an upper limit (penalty), it will only affect the machine score when the utilization goes above this limit. The plot displays this by showing a dot on 0 when it has no affect. Figure~\ref{fig:wl2_tanh_eucl} displays the same data as in Figure~\ref{fig:wl2_tanh_avg}, except that the overall machine score is computed using the hyperbolic tangent function and $\ell_2$-norm.

\begin{figure*}[!htb]
\centering
\includegraphics[scale=0.35]{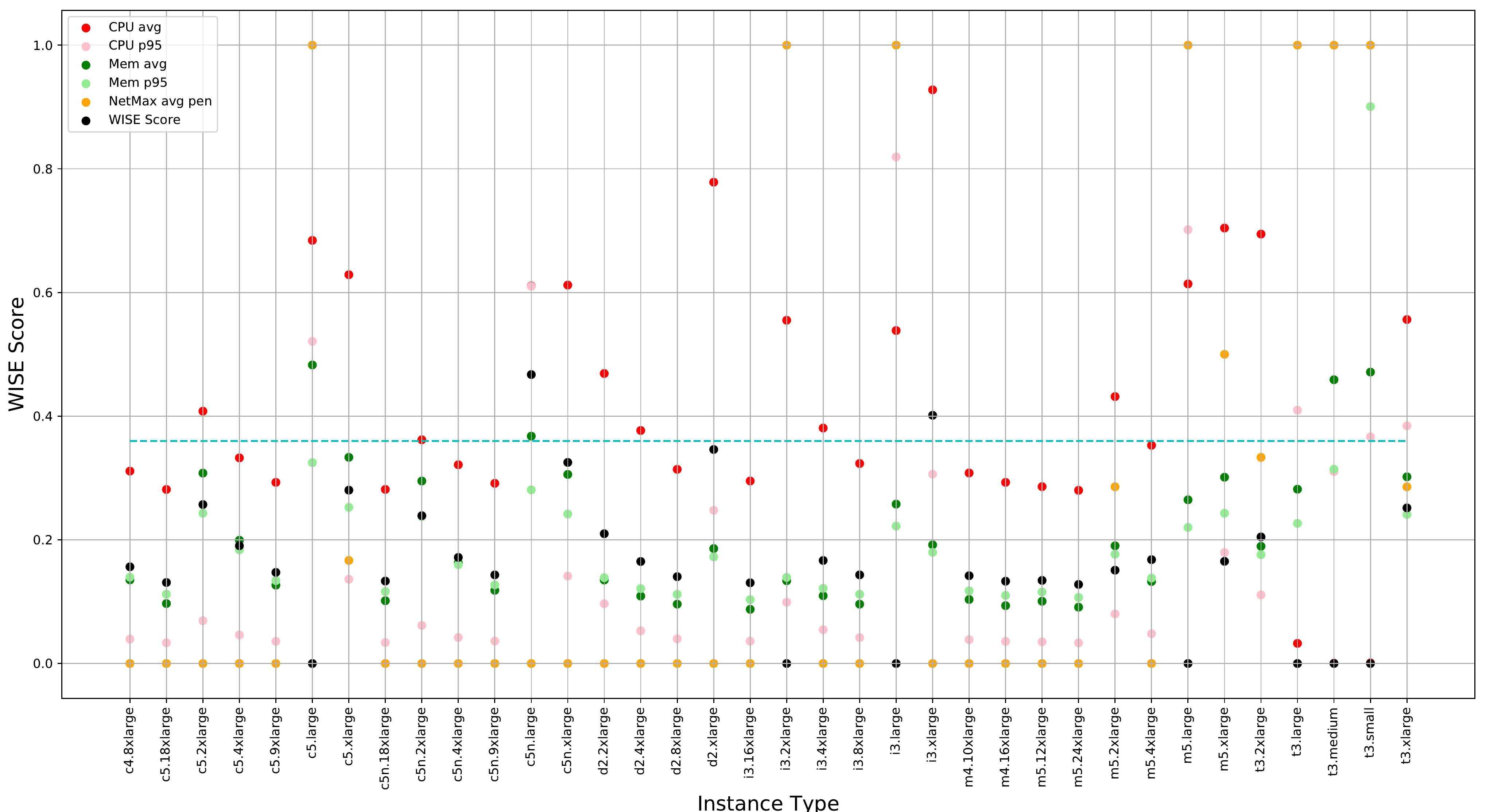}
\caption{WISE scores using the exponentially monotone function and $\ell_1$-norm for Streaming Workload.}
\label{fig:wl2_exp_avg}
\end{figure*}

\begin{figure*}[!htb]
\centering
\includegraphics[scale=0.35]{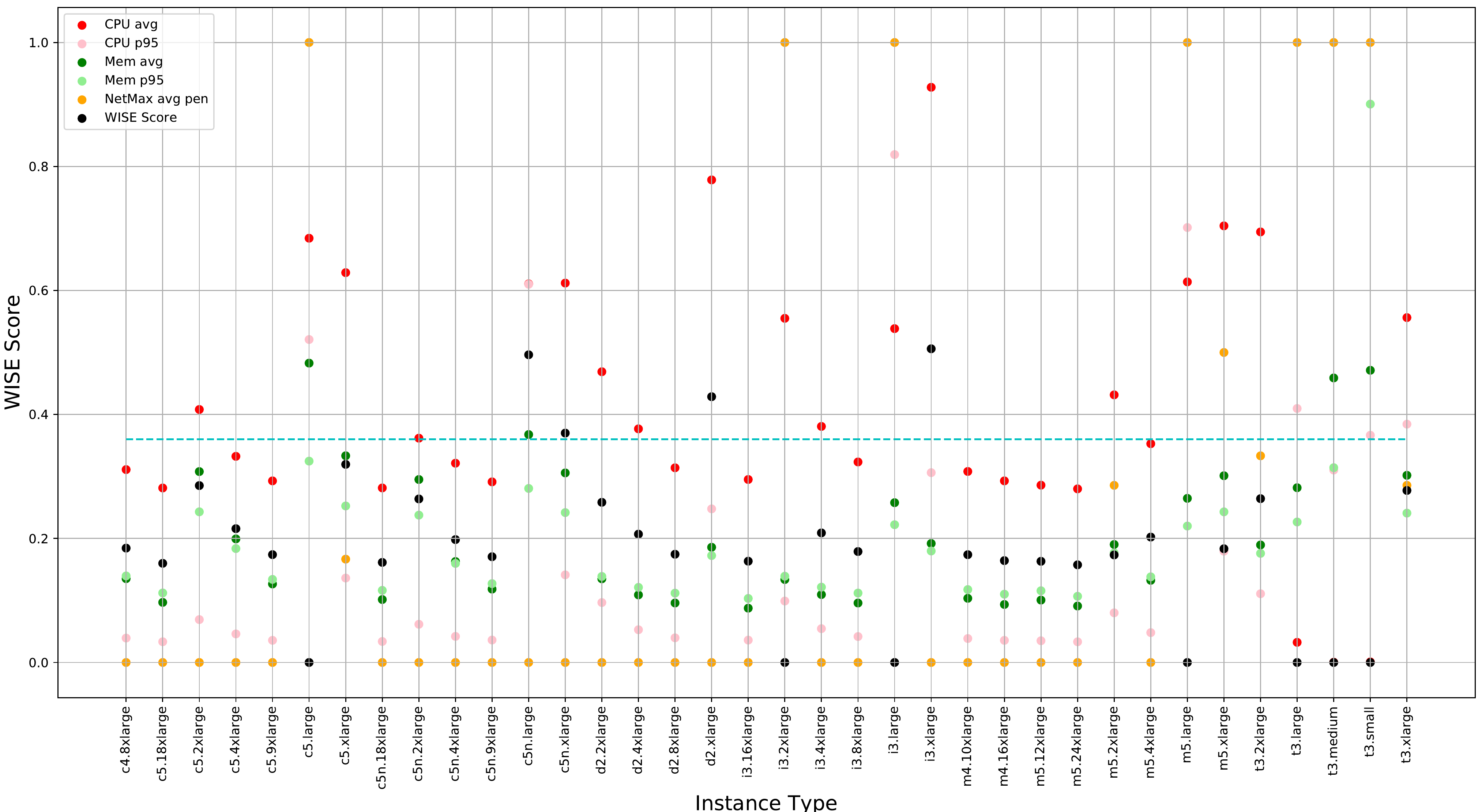}
\caption{WISE scores using the exponentially monotone function and $\ell_2$-norm for Streaming Workload.}
\label{fig:wl2_exp_eucl}
\end{figure*}

Figure \ref{fig:wl2_exp_avg} displays the WISE score and resource scores for a Streaming workload, except the normalization function using the exponentially monotone function and $\ell_1$-norm, with the best score having a value of 1. The area above the light blue dotted lines indicates values that fall within the described acceptable ranges. Any score that falls below this region indicates over- or under-utilization from the acceptable ranges. A machine score can still fall within the acceptable range while having a resource that falls out of range. Figure~\ref{fig:wl2_exp_eucl} displays the same data as in Figure~\ref{fig:wl2_exp_avg}, except that the overall machine score is computed using the exponentially monotone function and $\ell_2$-norm.

\section{Validation}
In this section, we describe and present results for a unique validation method that uses performance data from the benchmarks to validate the efficacy of the WISE score to determine well performing workload/machine combinations.

\subsection{Method}

We use two benchmarks to validate the WISE score 1) MongoDB, a CPU intensive benchmark. Configured to run steadily over time, 2) MongoDB, configured to run bursty with spikes of activity and 3) Streaming workload, a networking intensive benchmark. We first determine the optimal instances by using the performance metrics that are generated by running a specific workload on many different configurations of virtual machines. Some of the performance metrics used are duration, latency and throughput. It is important to note that during this phase of validation the WISE score does not come into play and neither does the utilization data. Optimal virtual machines are determined solely on the performance metrics of the benchmarks. We then compare the optimal virtual machine selected by using the performance metrics with the optimal virtual machines generated by using the WISE score using utilization data to determine if indeed, the WISE score is a good indicator of workload/machine performance.

\subsubsection{Performance Metrics}
Using the performance metrics generated for each workload/machine combination by the benchmark, the following
criteria is used to determine optimal instances for a specified workload. First, using the statistical inter-quartile range outlier detection method, remove any data points that have a latency or duration of greater than $Q3+1.5*IQR$. This removes any instances that are taking longer to execute with respect to duration and latency, they are outside of the range of most other instances. Second, the remaining instances are sorted by usage cost and the instances that fall within 3 times the usage cost of the cheapest instance (after part one) are selected. The first part removes instances that are under-provisioned and the second part removes instances that are over-provisioned. Resulting in instances that are capable of handling the load without being over-provisioned.

\subsubsection{WISE Score}
During this phase, the WISE score is calculated for each workload/machine combination. So for every instance that the benchmark is run on, we take the utilization data generated and get a WISE score. None of the performance metrics used in the first phase are used to calculate the WISE score. The WISE score only uses utilization data, no specific metrics from the workload such as latency, duration are used in the WISE Score. We use the Streaming workload to validate the WISE score, by first coming up with optimal instances using the performance metrics generated by running the workload on many instances such as latency and duration. We then compare the optimal instances generated from the performance metrics with the optimal instances generated by the WISE score using various standard evaluation metrics.

\subsubsection{WISE Score - Quality Tenets}	
The quality tenants for the proposed WISE score can be summarized as follows:
\begin{enumerate}
\item Instances with a WISE score above threshold should be able to run the workload with acceptable performance. The precision metric is used as an indicator of acceptable performance, indicating percentage of returned machines the run optimally according to performance metrics.
\item How to account for acceptable performing instances that are under-utilized? Although a machine has good performance metrics, it is possible that it is over provisioned for that specific workload. Therefore, from the list acceptable performing instances, only return those that are within 2 times the price of the cheapest instance.
\item How to account for acceptable performing instances that are over-utilized? Although a machine has good performance metrics, it is possible that it is over-utilized for the specific task, e.g. running very close to its capacity limits, but have not yet a resource wall yet. Use a tighter outlier cut-off point on the performance metrics. By using a tighter outlier cut-off point we will eliminate any instance types that may be close to that resource utilization wall. By looking at only the performance metrics and not the utilization data, we are not fully able to determine this, however, the solutions mentioned above mitigate some of the issues.
\item Use a ranking metric to determine how well the WISE score rankings compare to the ranked performance based list sorted by price. Note that this is not perfect as the WISE score does not take price into account, and does not rank a cheaper instance higher. It will rank higher, an instance that runs within acceptable utilization ranges. For example in the performance metrics a cheap instance will be ranked higher even if it is over-utilized.
\item How well does the WISE score identify well performing instances at reasonable prices? Reasonable prices here would be defined as in a range of 2/3 times the cheapest well performing instance. We use recall to determine how many of these.
\end{enumerate}

\subsubsection{Results}

Using the performance metrics and methods described above, for each workload we get a list of optimal performing virtual machines. The list derived by using the WISE scores will be compared to this optimal list. The precision and recall metrics are used along with a rank based metric, rank biased overlap \cite{webber2010similarity} to evaluate the ordering of WISE scores. We used a standard cutoff of $0.36$ for the exponential function and $0.76$ for the tangent function. Results show that the WISE score consistently identifies optimal instance types for a workload, see Table\ref{Tab:Validation_Results} and the precision metric. Recall shows how many of the optimal instance types is the WISE score able to identify, the exponential function with the $\ell_2$ performs best with this metric. The ranking metric shows how the WISE score orders the instance types by score. All of these metrics can be tweaked by changing the acceptable threshold parameter.

\begin{table*}[!htb]
	\rowcolors{1}{}{lightblue}
	\centering
	\caption{WISE Score validation results using benchmark performance metrics. Boldface numbers indicate the best evaluation metrics.}
	\medskip
	\begin{tabular}{l*{6}{c}}
		\toprule
		Benchmark & Function & Norm & Precision (\%) & Recall (\%) & Ranking\\
		\midrule
		MongoDB (steady)   & Tanh & $\ell_1$ & \textbf{1.0} & 0.667 & 0.699  \\
			               &      & $\ell_2$ & \textbf{1.0} & 0.333 & 0.51  \\
		                   & Exp. & $\ell_1$ & \textbf{1.0} & 0.333 & 0.51  \\
		                   &      & $\ell_2$ & 0.858 & \textbf{1.0} & \textbf{0.897}  \\
		MongoDB (bursty)   & Tanh & $\ell_1$ & \textbf{1.0} & 0.778 & 0.806  \\
				           &      & $\ell_2$ & \textbf{1.0} & 0.556 & 0.564  \\
				           & Exp. & $\ell_1$ & \textbf{1.0} & 0.556 & 0.564  \\
				           &      & $\ell_2$ & 0.889 & \textbf{0.889} & \textbf{0.849}  \\
	    Streaming          & Tanh & $\ell_1$ & \textbf{1.0} & \textbf{0.667} & 0.704  \\
	    		           &      & $\ell_2$ & \textbf{1.0} & 0.333 & \textbf{0.842}  \\
	    		           & Exp. & $\ell_1$ & \textbf{1.0} & \textbf{0.667} & 0.704  \\
	    		           &      & $\ell_2$ & 0.4 & 0.667 & 0575  \\
        \bottomrule
	\end{tabular}
	\label{Tab:Validation_Results}
\end{table*}

\section{Discussion}
The WISE framework may suffer from the curse of dimensionality if irrelevant dimensions (resources) are added. As less important resources are added, WISE can become inefficient as the more important resources become diluted by the less important ones. To circumvent this issue, only necessary resources and aggregations that have a certain amount of information gain should be used, as the impact of one resource is diminished by the number of resources given,  his can also be controlled by the weight factor. In essence it is important to add resources and aggregations that add value in determining performance. Moreover, it would be interesting to do this automatically by computing the information gain of each attribute and using only the most informative ones. This could be accomplished by first discarding all attributes whose information gains are below a pre-defined threshold and then measuring distance only in the reduced space.

We have designed the WISE score to be very flexible in its configuration possibilities. We have observed different configuration use cases and tolerances for over- and under-provisioning. Also, certain types of machines such as scientific supercomputers or GPU intensive computers can have different configuration levels. It is indeed a challenging issue to find those optimal values, and the existing baselines vary based on the use cases. There are various methods to set these levels, including through observation and expert knowledge, as well as understanding the needs of a workload, awareness of a business needs and cycles. By contrast, the proposed method selects a group of optimally running machines and trains the WISE model to output optimal scores for these machines. That configured WISE model can then be used to get tuned WISE scores for other machine/workload combinations.

\section{Conclusion}
In this paper, we proposed a novel approach for scoring a workload/machine combination representing the fitness of a machine running a particular workload. The WISE framework is is powerful in that it produces an index between 0 and 1, indicating the level of fitness for the workload/machine combination. It is flexible in that customers can define individualized targets and ranges to suit their needs and then use the WISE score to test their fleet of machines. WISE also allows for general definitions of proper machine running levels to very sophisticated resource definitions. Experimental results showed the efficacy of the proposed framework on two distinct workload configurations. Validation results showed that the WISE score was able to deliver optimal instance types on three different benchmark configurations. For future work, we plan to learn the resources' weights given some ground truth data and then use the WISE framework with the learned weights to compute the WISE scores.			
\bibliographystyle{ieeetr}
\bibliography{references}

\end{document}